# The formation of polymer-dopant aggregates as a possible origin of limited doping efficiency at high dopant concentration

Julie Euvrard [a], Amélie Revaux [a,*], Pierre-Alain Bayle [b], Michel Bardet [b], Dominique Vuillaume [c], and Antoine Kahn [d]

[a] Univ. Grenoble Alpes, CEA-LITEN, Grenoble, 38000, France

[b] CEA, INAC, SCIB UMR-E 3 CEA-UJF, Laboratoire de Résonance Magnétique, Grenoble, 38000, France

[c] IEMN, CNRS, Univ. Lille, Villeneuve d'Ascq, 59652, France

[d] Dept. of Electrical Engineering, Princeton University, Princeton, NJ, 08544, USA

The polymer Poly[(4,8-bis-(2-ethylhexyloxy)-benzo(1,2-b:4,5-b')dithiophene)-2,6-diyl-alt-(4-(2-ethylhexanoyl)-thieno[3,4-b]thiophene-)-2-6-diyl)] (PBDTTT-c) p-doped with the molecular dopant tris[1-(trifluoroethanoyl)-2-(trifluoromethyl)ethane-1,2-dithiolene] (Mo(tfd-COCF$_3$)$_3$) exhibits a decline in transport properties at high doping concentrations, which limits the performance attainable through organic semiconductor doping. Scanning Electron Microscopy is used to correlate the evolution of hole conductivity and hopping transport activation energy with the formation of aggregates in the layer. Transmission Electron Microscopy with energy-dispersive X-ray analysis along with liquid-state Nuclear Magnetic Resonance experiments are carried out to determine the composition of the aggregates. This study offers an explanation to the limited efficiency of doping at high dopant concentrations and reinforces the need to increase doping efficiency in order to be able to reduce the dopant concentration and not negatively affect conductivity.

1. Introduction

Organic printed electronics is a promising route for the next generation of electronic devices. With large area scalability, compatibility with flexible substrates, and low temperature processability, printed electronics offers an interesting alternative to conventional silicon-based electronics and aims at targeting new applications [1]. In order to achieve better device performance, transport properties in the organic semiconductor and at the semiconductor-metal electrode interfaces must be improved [2]. In that regard, the development of controlled, efficient and stable doping is known to be highly beneficial [3]. Doping is used to increase carrier mobility and semiconductor conductivity through the addition of free carriers [4,5]. In addition, spatially localized doping can be used to realize ohmic contacts at semiconductor-metal interfaces [6–8].

---

* Corresponding author. Tel.: +33 438784593.
E-mail address: amelie.revaux@cea.fr (A. Revaux).

Molecular p-dopants as tetracyano-quinodimethane (TCNQ) derivatives or dicyanodichloro-quinone (DDQ) have been introduced in various polymers, showing effective p-doping ability [9,10]. Increased conductivity, carrier mobility and hole density have been reported for multiple polymer-dopant mixtures [7,11]. However, to reach sufficient doping impact, high concentrations of molecular dopants (several % in molar ratio) need to be added to the polymer matrix, in contrast to inorganic semiconductors where concentrations of the order of $10^{-6}$-$10^{-3}$ dopant per semiconductor atom are required [12]. High concentrations of organic molecules can lead to a degradation of the transport properties through the formation of defects [13,14]. The conductivity saturation or decline above a certain dopant concentration threshold has been widely reported in the literature [15–17]. As organic electronic devices require efficiently doped layers, it is crucial to understand the origins of the limited doping efficiency in order to unlock the boundaries of organic semiconductor doping.

In this study, we use a soluble derivative of Mo(tfd)$_3$, i.e. tris[1-(trifluoroethanoyl)-2-(trifluoromethyl)ethane-1,2-dithiolene] (Mo(tfd-COCF$_3$)$_3$), to p-dope the polymer Poly[(4,8-bis-(2-ethylhexyloxy)-benzo(1,2-b:4,5-b')dithiophene)-2,6-diyl-alt-(4-(2-ethylhexanoyl)-thieno[3,4-b]thiophene-)-2-6-diyl)] (PBDTTT-c). Electrical characterizations are carried out as a first step to determine the evolution of the hole conductivity and hopping transport activation energy with the doping concentration, highlighting the lower doping efficiency at high doping concentration. Scanning Electron Microscopy (SEM) is used to correlate the electrical characteristics with changes in the layer morphology. Finally, Transmission Electron Microscopy (TEM) with energy-dispersive X-ray (EDX) analysis and liquid state Nuclear Magnetic Resonance (NMR) are carried out to further understand the morphology evolution.

## 2. Experimental methodology

PBDTTT-c, purchased from Solarmer Materials, and the dopant Mo(tfd-COCF$_3$)$_3$ are dissolved separately in ortho-xylene and stirred for 12 hours at 45°C. To obtain the solution of doped polymer, the appropriate volume of dopant solution is added to the polymer solution. The resulting mixture is stirred at 45°C for 4 hours. The concentration of dopants in the polymer matrix is determined in molar ratio (MR), which corresponds to the number of dopant molecules added per monomer of PBDTTT-c. The solution of pure or doped polymer is deposited using spin-coating technique and annealed.

To study the lateral transport of holes, we used a hole-only structure with the polymer layer deposited on inter-digitated gold electrodes (100 nm) evaporated on a glass substrate. A thin layer of titanium (10 nm) is evaporated prior to gold to ensure a good adhesion to the quartz substrate. This device is processed in a glovebox and the polymer layer is annealed at 150°C for 10 min. The thickness of each layers is measured using an Atomic Force

Microscope (AFM) in a nitrogen atmosphere. After processing, the device is transferred to a vacuum chamber (base pressure of ) for measurements. The temperature is varied from 200 to 390 K in steps of 5 K using a closed-cycle He refrigerator combined with a heater. Current density-voltage (J(V)) measurements are carried out from -50 V to +50 V in steps of 2 V using a Keithley 2400 source meter.

For SEM analysis, the pure and doped polymer solutions are stirred overnight at 45°C and deposited on glass substrates as explained previously. A few monolayers of platinum are deposited on the pure or doped layer to avoid charge build-up in the substrate. Measurements are carried out with a SEM Leo 1530 from Zeiss at an accelerating voltage of 3 kV. Secondary electrons are used for this analysis and the working distance varies from 2.7 to 3.5 mm. A FEI Tecnai Osiris TEM microscope, operated at an accelerating voltage of 200 kV, was used for TEM analysis. A solution of Mo(tfd-COCF$_3$)$_3$ doped PBDTTT-c at 5% MR is deposited on a metal grid. Finally, NMR measurements were carried out in liquid state and solid state using respectively 400 MHz and 500 MHz Bruker NMR spectrometers. Solutions of pure and doped polymer at 5% MR were analyzed.

## 3. Results and Discussion

### 3.1. Electrical characterization

J(V) measurements were carried out as a function of temperature on PBDTTT-c films doped at 0.5% to 8% MR. **Fig. 1** (a) shows the current density with respect to the electric field in a logarithmic scale at 300 K. Only 4 doping concentrations are shown for the sake of clarity, although the evolution of the current follows the trend suggested by the arrows for all cases. The graph highlights the rise in current density by almost 3 orders of magnitude for doping concentrations varying from 0.5% to 2.5% MR. Then, the current density decreases by one order of magnitude between 2.5% and 8% MR.

The resistance $R$ is extracted for each doping concentration from the J(V) characteristics in the ohmic regime (below ) and used to calculate the conductivity ($\sigma$). The error of the extracted resistance and the uncertainty of the layer thickness estimated around 10 nm from AFM measurements are taken into account to determine the uncertainty associated with σ. A contact resistance of  has been extracted at the gold electrode using Transmission Line Measurement (TLM) on a sample of pure polymer, while the resistance of the bulk is measured around  (supplementary information, **FIG. S1 (a)**). As an ohmic contact is formed with doping [18], we expect the contact resistance to decrease. Moreover, a TLM analysis on a 5% MR doped sample exhibits a contact resistance 4 orders of magnitude lower than the bulk resistance (supplementary information, **FIG. S1 (b)**). Therefore, the contact resistance can be neglected for the pure and doped polymer. The evolution of the PBDTTT-c conductivity at 300 K with Mo(tfd-COCF$_3$)$_3$ doping is given in **Fig. 1** (b). From 0.5% to 2.5% MR, σ follows a superlinear increase with increasing doping concentration, reaching ~. This superlinear increase has been attributed to trap filling and has been observed in several organic semiconductor-dopant mixtures [3,7,14]. The discrepancy of the data point at 1% MR might be due to inhomogeneity in the layer thickness. For doping concentrations above 2.5% MR, a slight decrease is observed with a power law of -0.5. This decline in σ at higher doping concentrations hinders the potential improvement of the transport properties required for organic semiconductors.

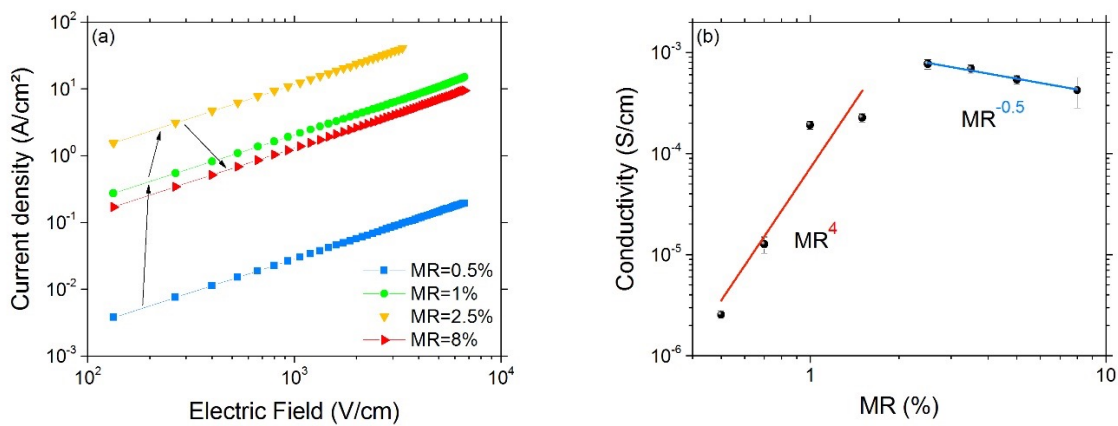

**Fig. 1.** (a) Current density with respect to the electric field in a logarithmic scale for 4 doping concentration at 300 K. (b) Conductivity of PBDTTT-c with respect to Mo(tfd-COCF$_3$)$_3$ molar ratio in a logarithmic scale at 300 K.

σ was also extracted at each concentration for temperatures varying between 200 and 390 K. **Fig. 2** (a) shows σ vs. 1000/T for the 8 doping concentrations studied. The linear behavior of the conductivity with 1/T in a semi-logarithmic scale is in agreement with hopping transport, a classical transport mechanism in organic semiconductors. When charge carriers move via hopping between localized states, can be described by an Arrhenius law [11,19]:

with the activation energy of hopping transport, a pre-factor, the Boltzmann constant and the temperature. Therefore, hole conductivity measurements at variable temperatures leads to the determination of hopping transport activation energy for holes. In this analysis we choose to neglect the temperature effect on the dopant ionization efficiency. This value has been measured for a concentration of 1% MR at (supplementary information, **Fig. S2**), an energy one order of magnitude lower than the hopping transport activation energy obtained for pure and doped PBDTTT-c.

**Fig. 2** (b) shows the evolution of the hole hopping transport activation energy vs. the doping concentration. decreases with doping up to 2.5% MR, reaching 240 meV. A slight deviation from the fit is observed at 1% MR as identified in the evolution of the conductivity. Above the threshold of 2.5% MR, the decrease of the activation energy is much lower and reaches 230 meV at 8% MR. The reduction of the activation energy is attributed to a reduction of the effect of traps on the transport of carriers, as the Fermi level shifts toward the transport level [14]. The initial decrease of observed here up to 2.5% MR for Mo(tfd-COCF$_3$)$_3$ doped PBDTTT-c and for several other polymer-dopant blends reported in the literature, is therefore associated with the gradual filling of trap states above the HOMO level [11,20,21], consistent with the superlinear increase in σ.

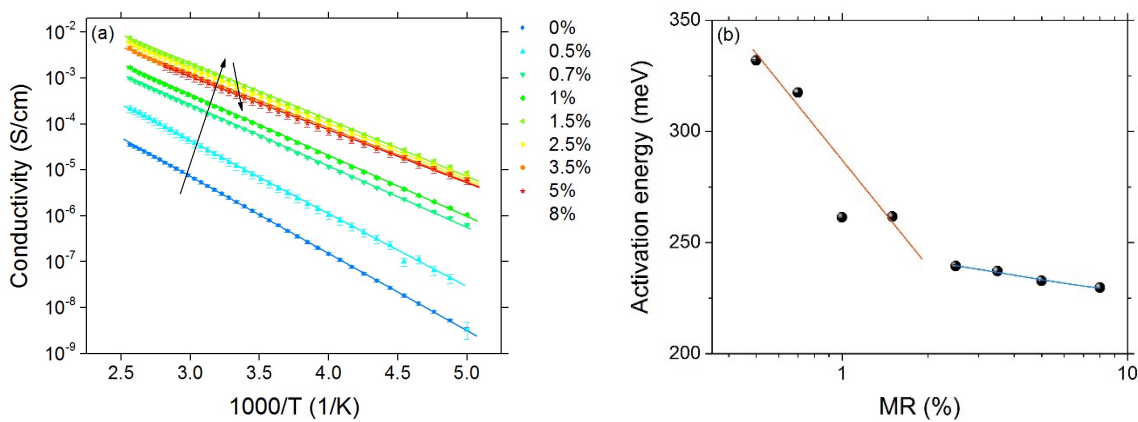

**Fig. 2.** (a) Arrhenius plot of conductivity for Mo(tfd-COCF$_3$)$_3$ doped PBDTTT-c with respect to the inverse of temperature in a semi-logarithmic scale and (b) corresponding hole hopping transport activation energy with respect to the dopant molar ratio.

The σ and data presented above highlight two different regimes. For doping concentrations below 2.5% MR, the evolution of both parameters can be explained by trap filling [3,7,14]. Once all traps are filled, σ is expected to increase linearly and will decrease continuously as the Fermi level shifts toward the polymer HOMO. However, we observe a slight decline in σ and the saturation of above 2.5% MR, which is usually related to additional trap states created in the polymer matrix upon dopant addition [14,16]. These energy levels are often attributed to strong Coulomb interaction with the ionized dopant [22], changes in the polymer morphology [23] or formation of aggregates [15,24]. It has also been observed that the Fermi level shift in organic semiconductor doping is limited and saturates several hundreds of meV above the polymer HOMO [25,26]. However, the origin of such saturation is still under discussion. In this study we have chosen to analyze the morphology evolution of the PBDTTT-c layer with the concentration of Mo(tfd-COCF$_3$)$_3$ and observe potential aggregates of dopants or polymer-dopant, which would explain the electrical behaviors observed above 2.5% MR.

### 3.2. Morphology evolution

SEM images were taken on pure PBDTTT-c and Mo(tfd-COCF$_3$)$_3$ doped PBDTTT-c with 7 doping concentrations varying from 0.5% to 6% MR and shown in **Fig. 3** (a) to (h). From 0% to 1% MR, the layer is uniform and no particular modification can be observed on the SEM images. At a doping concentration of 2% MR, brighter nanoparticles can be identified on the layer surface. The increase of brightness can be due to an increase of the atomic number Z of the components. As the molybdenum atomic number ($Z_{Mo}$=42) is higher than the atomic number of all other elements in the polymer ($Z_C$=6, $Z_O$=8, $Z_S$=16 and $Z_H$=1), an increase of Z would be consistent with the formation of aggregates containing the molybdenum complex.

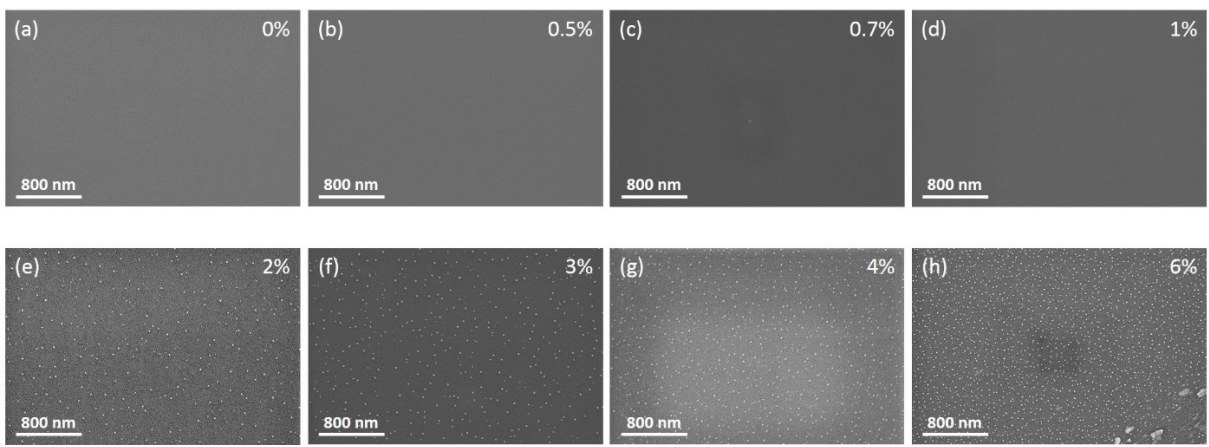

**Fig. 3.** Scanning Electron Microscope (SEM) images of pure PBDTTT-c (a) and doped PBDTTT-c at 0.5% MR (b), 0.7% MR (c), 1% MR (d), 2% MR (e), 3% MR (f), 4% MR (g) and 6% MR (g) with a magnitude of $10^5$ X.

**Fig. 4** (a) to (c) shows higher magnification SEM images for the doping concentrations 2%, 3% and 4% MR. The sizes of the particles are estimated around 20 nm in diameter. Increasing the doping concentration above 2% MR increases the density of aggregates visible on the surface of the layer but has no impact on their size.

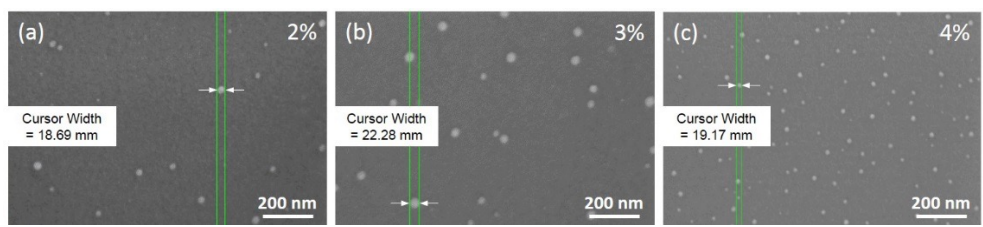

**Fig. 4.** Scanning Electron Microscope (SEM) images of Mo(tfd-COCF$_3$)$_3$ doped PBDTTT-c at 2% MR (a), 3% MR (b) and 4% MR (c) with a magnitude of $3.10^5$ X.

The formation of aggregates or the occurrence of phase segregation have been reported in the literature for several doped organic semiconductors. Deschler *et al.* [23] have reported the formation of dopant rich domains in 2,3,5,6-Tetrafluoro-7,7,8,8-tetracyanoquinodimethane (F$_4$-TCNQ) doped poly(3-hexylthiophene) (P3HT) and

Poly[2,1,3-benzothiadiazole-4,7-diyl[4,4-bis(2-ethylhexyl)-4H-cyclopenta[2,1-b:3,4-b']dithiophene-2,6-diyl]] (PCPDTBT). They correlated the phase separation with the conductivity saturation observed for both polymer-dopant blends. The aggregation of doped polymer domains has been identified in $F_4$-TCNQ doped Poly[2,5-bis(3-tetradecylthiophen-2-yl)thieno[3,2-b]thiophene] (PBTTT-$C_{14}$) by Cochran *et al.* [27]. In the case of aggregates formed in the PBDTTT-c:Mo(tfd-$COCF_3$)$_3$ layer above 2% MR, further analyses are required to determine whether aggregates are formed of pure dopant or polymer-dopant mixture.

### 3.3. Composition of the aggregates

TEM images were taken on a 5% doped sample and shown in **Fig. 5** (a). Aggregates with varying shapes and sizes are observed. We believe that the bigger aggregates are due to the formation of clusters with the smaller aggregates observed in SEM images. Admittedly, differences can occur in the size of the aggregates between SEM and TEM analyses as one experiment is carried out on annealed thin films while the other is conducted in solution. However, the composition should not be affected by the film formation and the annealing step as the polymer and dopant do not evaporate at the temperature used for annealing. Therefore, TEM analysis is a useful tool to determine the composition of the aggregates.

An EDX analysis is carried out on the cluster shown in **Fig. 5** (a). **Fig. 5** (b), (c) and (d) exhibit the map of fluorine, sulfur and oxygen respectively on the area scanned by the X-ray detector. The EDX spectrum is given in supplementary information (**Fig. S3.**). Sulfur and oxygen are found on the whole area, although the cluster exhibits a higher density of both components. Fluorine is mainly observed in the cluster, indicating that Mo(tfd-$COCF_3$)$_3$ molecules are probably mostly situated in the aggregates. However, we cannot determine whether the aggregates contain some PBDTTT-c molecules or dopant only, as no atom (except hydrogen) is specific to the polymer.

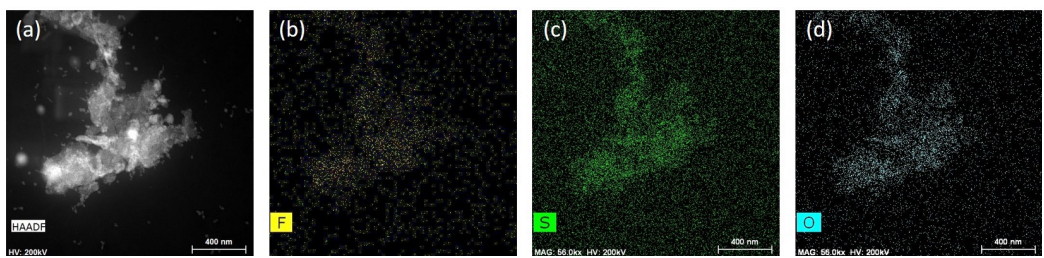

**Fig. 5.** TEM image of the 5% MR sample obtained with a magnification of 56kX (a) and the fluorine (b), sulfur (c) and oxygen (d) maps obtained by EDX.

In order to obtain more insights on the composition of the aggregates, we carried out $^{19}$F NMR analysis on pure and doped PBDTTT-c. Liquid state NMR has been chosen to obtain NMR spectra of pure dopant using small quantities of dopant powder. Prior to the analysis in liquid state $^{19}$F NMR, $^{13}$C NMR on doped polymer was carried out in liquid and solid state using Magic Angle Spinning (MAS) to ensure that the polymer-dopant interaction is similar in solution and solid thin film. The $^{13}$C spectra of pure and doped PBDTTT-c conducted in solid and liquid state are given in supplementary information (**Fig. S4.** and **Fig. S5.**). In solid as in liquid state NMR, the addition of dopants in the polymer matrix mainly modifies the $^{13}$C spectrum between 110 and 160 ppm, with the attenuation of the peaks corresponding to the carbon atoms contained in the thiophene rings. Such attenuation, usually due to a broadening of the peaks, can be explained by the paramagnetic influence of molybdenum leading to a faster spin-spin $T_2$ relaxation [28]. This evolution indicates that the molybdenum complex might be situated close to the thiophene rings. The comparison of solid and liquid state $^{13}$C NMR for pure and doped PBDTTT-c shows that the polymer-dopant interaction already takes place in solution. Therefore, NMR spectroscopy of pure dopant and doped polymer in liquid state is sufficient to determine whether all molecular dopants interact with the polymer. If the aggregates observed in SEM images are made of pure dopant, NMR spectroscopy should reveal the presence of Mo(tfd-COCF$_3$)$_3$ in its pristine form in the polymer-dopant mixture.

**Fig. 6** exhibits the liquid state $^{19}$F spectra of the pure Mo(tfd-COCF$_3$)$_3$ dopant and the doped polymer at 5% MR. For the pure dopant, two peaks are observed around -55 ppm and -73 ppm. The integration of both peaks leads to a ratio close to 1:1. Therefore, the two peaks are probably associated with the two types of carbon-fluorine bonds (green and blue in the chemical structure given as inset). The presence of the carbonyl group close to the carbon-fluorine bonds colored in green might lead to a different chemical shift of $^{19}$F. When the dopant is mixed with the polymer, the two peaks combine to form one broad peak centered around -62 ppm. The remaining thin peaks visible in the pure dopant and the polymer-dopant mixture are probably due to impurities or ligands separated from the complex. To make sure that the broader peak obtained after addition of the dopant in the polymer matrix is not due to a probe noise, we measured the $^{19}$F spectra of the pure dopant and doped polymer under the exact same experimental conditions (see **Fig. S6.**), confirming the absence of peaks around -62 ppm in the mixture.

Although the structural changes of the dopant is not identified, the observed modification in the $^{19}$F spectrum is unambiguously assigned to the dopant interaction with the polymer. Indeed, if some molecular dopants had not interacted with the polymer host, we would expect small peaks around -55 ppm and -73 ppm corresponding to the dopant in its pristine form, which are not observed in the doped polymer. Therefore, we can conclude that all molecular dopants added to the polymer have reacted. It is then unlikely that aggregates of pure dopant are formed. As a result, we can conclude that the aggregates observed above 2% MR in SEM and TEM images are composed of polymer-dopant mixture.

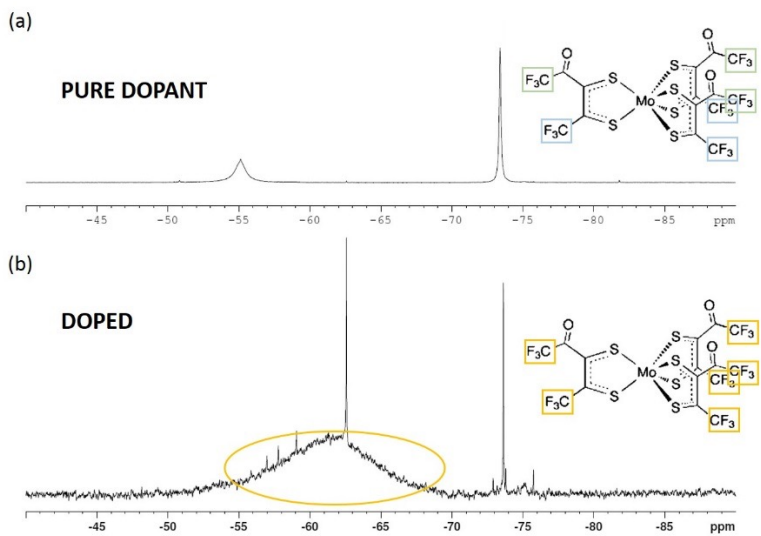

**Fig. 6.** Liquid-state $^{19}$F NMR spectra for pure Mo(tfd-COCF$_3$)$_3$ dopant (a) and 5% MR doped PBDTTT-c with Mo(tfd-COCF$_3$)$_3$ (b). The chemical structure of Mo(tfd-COCF$_3$)$_3$ is given as inset.

Polymer-dopant aggregates in the pure polymer matrix can lead to a decline in the transport properties as the layer is made of sparse highly doped clusters with a high mobility in a pure polymer phase, which exhibits a limited mobility. This separation between pure polymer phase and clusters of doped polymer (illustrated in **Fig. S7.**) can explain the hole conductivity decrease above 2.5% MR as well as the saturation of the hole hopping transport activation energy. However, the evolution of the transport properties can also be due to the formation of additional trap states or the Fermi level pinning around 2.5% MR [14,25]. Further analyses are required to determine whether the formation of aggregates fully explains the transport properties at high doping concentration.

As the formation of aggregates above 2% MR is consistent with the evolution of the electrical characteristics, it is likely that this separation between pure polymer and doped polymer phases plays a role in the transport properties decline. The deposition of molecular dopants on the polymer thin film, doping the semiconductor through diffusion in the layer as suggested by Guillain *et al.* [29], could prevent the formation of polymer-dopant aggregates. However, this technique based on the ability of the dopant to diffuse in the polymer matrix can induce stability issues. Further works are necessary to obtain a more homogeneous doped layer and push the boundaries of organic semiconductor doping performances.

## 4. Conclusion

A decline in the transport properties has been highlighted for Mo(tfd-COCF$_3$)$_3$ doped PBDTTT-c at high concentrations through hole conductivity and hopping transport activation energy measurements. As changes in morphology can explain a degradation of the carrier transport, SEM analysis was carried out for different doping concentrations, showing the formation of aggregates above 2% MR. According to TEM with EDX analyses and NMR experiments, the aggregates are composed of polymer-dopant mixture as all molecular dopants have reacted with the polymer host. The segregation of phases with presumably high and low carrier mobilities could explain the evolution of the transport properties above 2.5% MR. Therefore, to improve the performances of organic semiconductor doping, polymer-dopant aggregates need to be avoided or the doping efficiency should be increased in order to reduce the quantity of molecular dopants required.

## ACKNOWLEDGMENTS

A part of this work was carried out at Princeton University and financially supported by a grant of the National Science Foundation (DMR-1506097). We gratefully acknowledge the group of Prof. Seth Marder for providing the dopants. We thank Xin Lin and Andrew Higgins for help with various sample preparations, Simon Charlot for the MEB images and Nathalie Pelissier for the TEM analysis.

# The Formation of polymer-dopant aggregates as a possible origin of limited doping efficiency at high dopant concentration

Julie Euvrard [a], Amélie Revaux [a,*], Pierre-Alain Bayle [b], Michel Bardet [b], Dominique Vuillaume [c], and Antoine Kahn [d]

[a] Univ. Grenoble Alpes, CEA-LITEN, Grenoble, 38000, France

[b] CEA, INAC, SCIB UMR-E 3 CEA-UJF, Laboratoire de Résonance Magnétique, Grenoble, 38000, France

[c] IEMN, CNRS, Univ. Lille, Villeneuve d'Ascq, 59652, France

[d] Dept. of Electrical Engineering, Princeton University, Princeton, NJ, 08544, USA

**SUPPLEMENTARY INFORMATION**

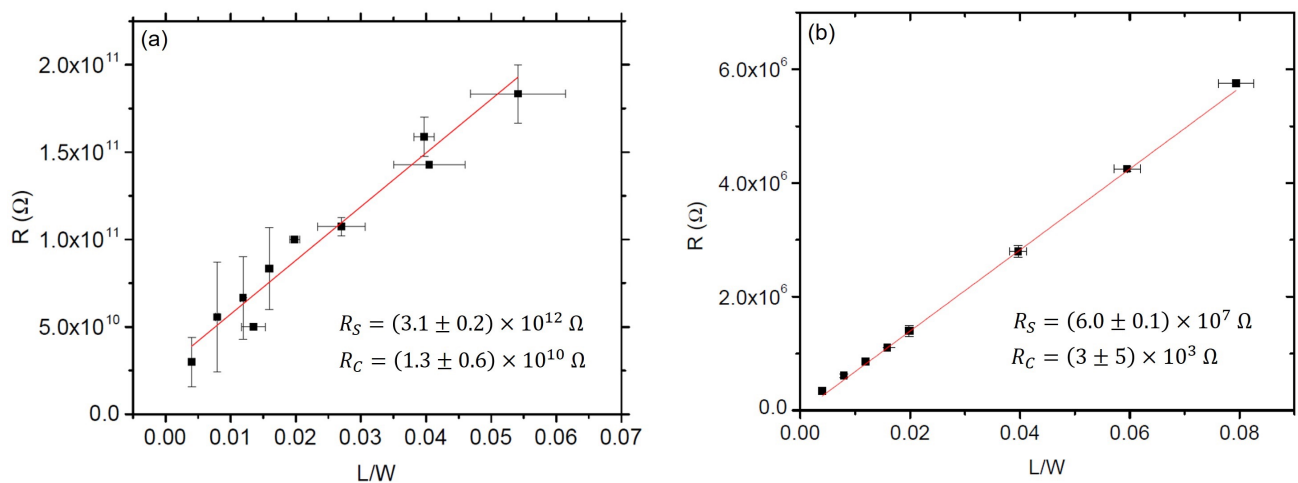

**Fig. S1:** Resistance with respect to L/W extracted from TLM analysis with a linear fit to extract the series resistance $R_S$ of the layer and the contact resistance $R_C$ with the gold electrode for pure polymer (a) and 5% MR doped polymer (b).

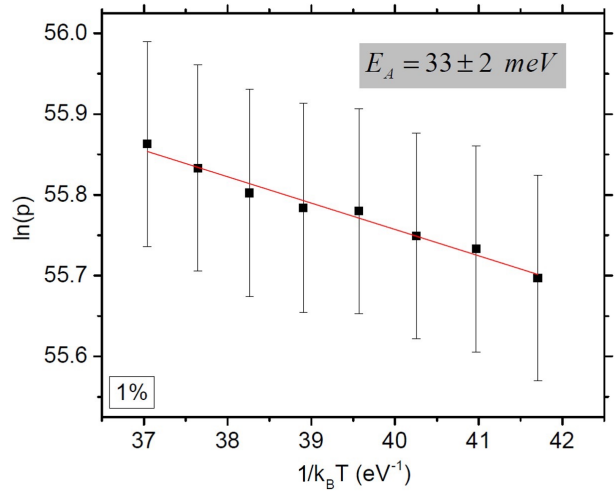

**Fig. S2.** Arrhenius plot of the doping activation with temperature for a concentration of 1% MR.

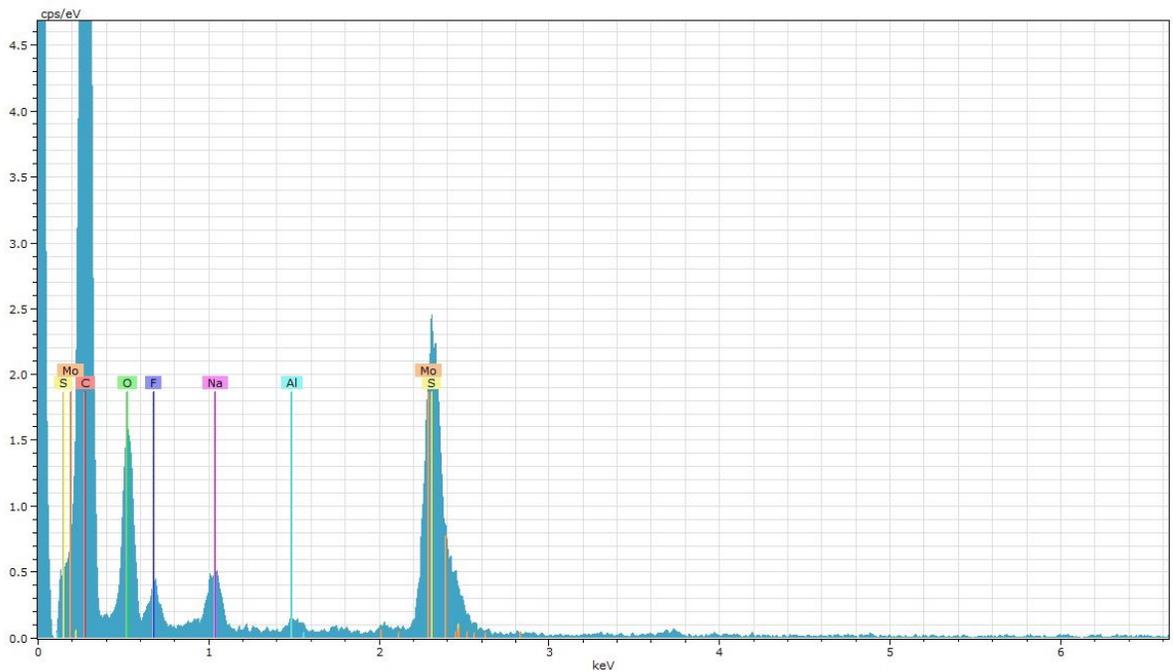

**Fig. S3.** EDX spectrum of the aggregates cluster analyzed for the 5% MR sample.

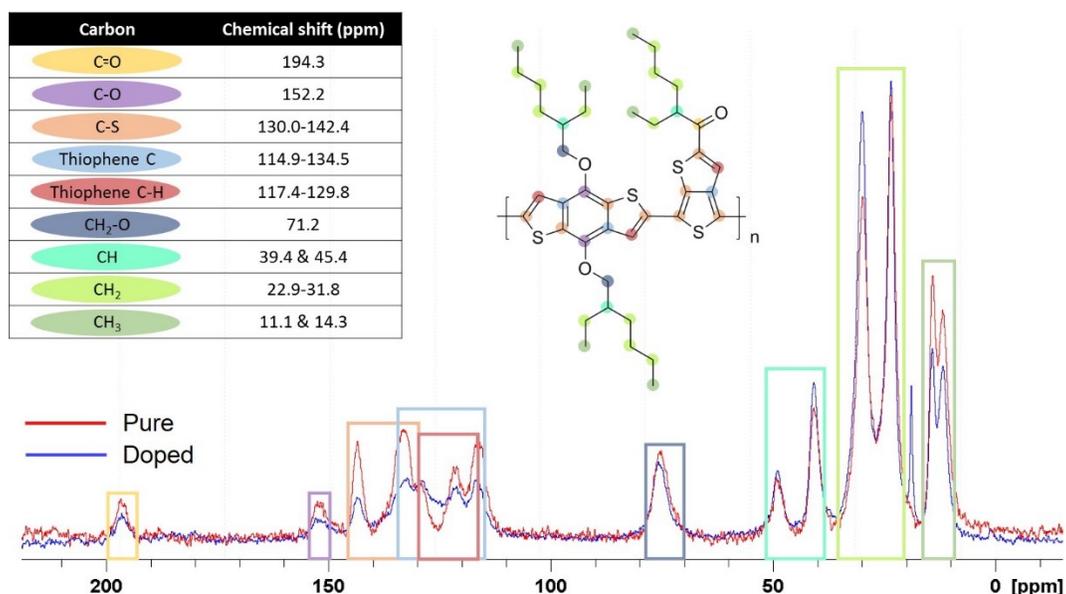

**Fig. S4**. $^{13}C$ spectra of solid-state MAS NMR for pure (red) and Mo(tfd-COCF$_3$)$_3$ doped PBDTTT-c (blue). The carbon chemical shifts simulated for a monomer of PBDTTT-c using nmrdb.org are given in the table as inset. Colors are used to highlight the peaks corresponding to the carbon atoms in the PBDTTT-c chemical structure given as inset.

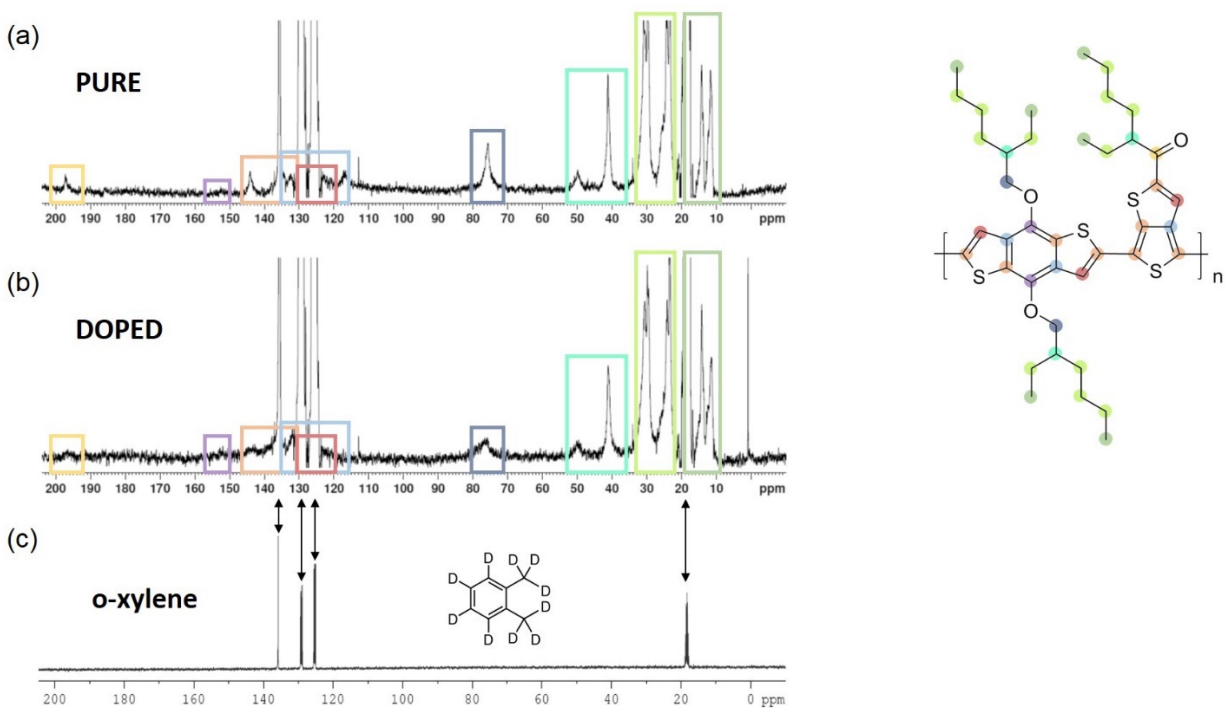

**Fig. S5**. $^{13}C$ spectra of liquid-state NMR for pure PBDTTT-c (a), Mo(tfd-COCF$_3$)$_3$ doped PBDTTT-c (b) and deuterated o-xylene (c). Colors are used to highlight the peaks corresponding to the carbon atoms in the PBDTTTT-c chemical structure given as inset.

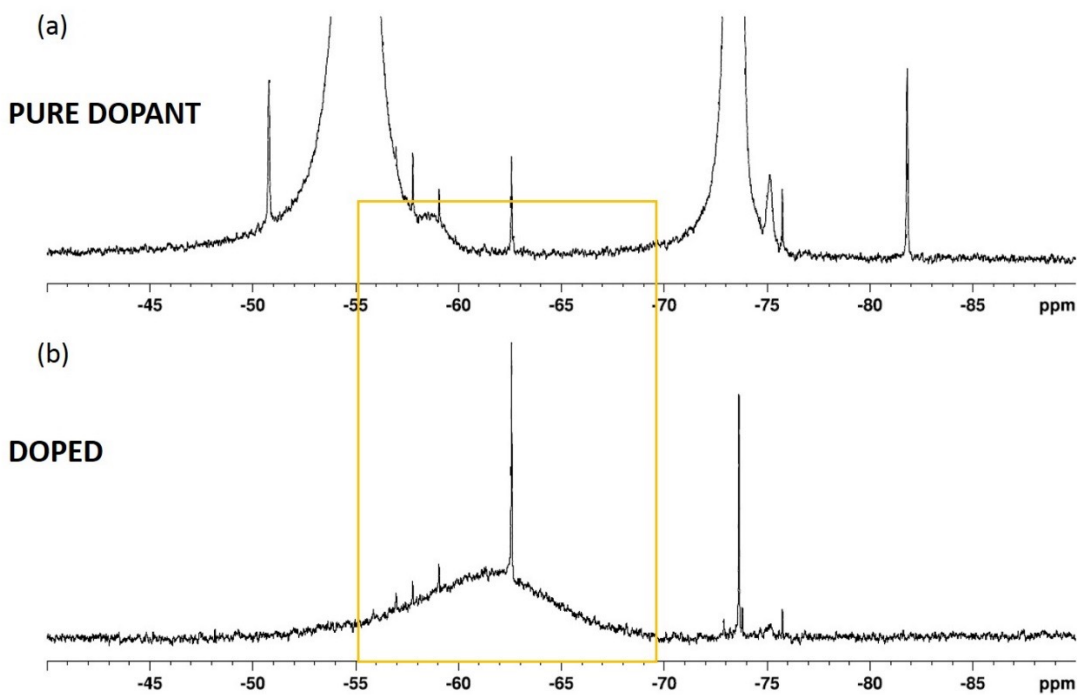

**Fig. S6.** $^{19}F$ spectra of liquid-state NMR for pure Mo(tfd-COCF$_3$)$_3$ dopant (a) and 5% MR doped PBDTTT-c with Mo(tfd-COCF$_3$)$_3$ (b) carried out with the exact same experimental conditions.

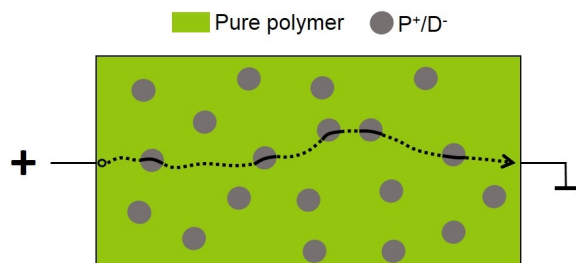

**Fig. S7.** Schematic of the PBDTTT-c layer (in green) with aggregates of Mo(tfd-COCF$_3$)$_3$ doped PBDTTT-c (in gray). The morphology impact on hole transport is illustrated with dashed and solid lines in pure (low mobility) and doped (high mobility) areas respectively.